\documentstyle[manuscript,aps,prbbib]{revtex}

\tighten
\newcommand{\be}{\begin{equation}}
\newcommand{\ee}{\end{equation}}
\newcommand{\bea}{\begin{eqnarray}}
\newcommand{\eea}{\end{eqnarray}}
\newcommand{\bd}{\begin{displaymath}}
\newcommand{\ed}{\end{displaymath}}
\newcommand{\bi}{\begin{itemize}}
\newcommand{\ei}{\end{itemize}}
\newcommand{\bc}{\begin{center}}
\newcommand{\ec}{\end{center}}
\newcommand{\bfl}{\begin{flushleft}}
\newcommand{\efl}{\end{flushleft}}
\newcommand{\bfr}{\begin{flushright}}
\newcommand{\efr}{\end{flushright}}
\newcommand{\f}{\frac}

\def\bk{{\bf k}} \def\bq{{\bf q}} \def\bp{{\bf p}} 
  
\def\ra{\rightarrow}
\def\6{\partial} \def\a{\alpha} 
  \def\ve{\varepsilon}

\def\o{\omega}  
  \def\S{\Sigma}

\def\={\!\!\!&=&\!\!\!}
\def\+{\!\!\!&&\!\!\!+~}
\def\-{\!\!\!&&\!\!\!-~}

\begin{document}

\title{Non-Fermi behavior of the disorder electronic system}
\author{C. P. Moca, I. Tifrea and M. Crisan}
\address{Department of Theoretical Physics\\
University of Cluj, 3400 Cluj, Romania}
\maketitle

\begin{abstract}
We showed that an electronic system with weak disorder
considered in the finite-charge infinite U Hubbard model can present
a non-Fermi behavior. The imaginary pert of the self-energy has been
calculated and a linear temperature dependence was obtained. This result
is in agreement with the non-Fermi behavior observed at the insulator-metal
crossover in La$_{2-x}$Sr$_x$CuO$_4$.
\end{abstract}

\section{Introduction}

The interplay between strongly correlated electron systems and disorder is
still an open question. The most efficient method is to start from the weak
disorder in the electronic system and to consider correlations using the
renormalization group (RG) method \cite{1,2,3}. The main result of this
method is prediction of the metal-insulator transition when the symmetry was
broken by the interaction with impurities.

The problem was studied using infinite-U Hubbard model and the t-J model
\cite{4,5,6}. The infinite dimension approach introduced by Metzner and
Vollhardt \cite{7} has been applied by many authors \cite{8,9,10} (for a
complete discussion see Ref. \onlinecite{10}) and using RG method  Si and
Kotliar \cite{11} showed that in an extended Hubbard model the disorder can
induce a non-Fermi behavior.

In this paper we will show that the weak disorder in the finite-charge
infinite-U Hubbard model can induced a non-Fermi behavior for a
two-dimensional (2D) electronic system close to the metal-insulator
transition. The paper is organized as follows. In Sect. II we present
the model. The self-energy of the electronic system is calculated in
Sect. III and we show that linear temperature dependence may appear,
which show a typical non-Fermi behavior. The relevance of our results
for the explanation of the experimental results will be discussed in
Sect. IV.

\section{Model}
We consider an electronic system with weak disorder in the finite-charge
infinite-U Hubbard  model. The charge susceptibility has been calculated
in \cite{6} as
\be
\chi_c(\bq,\o)=\f{N(0) N D q^2}
{2\left[Dq^2A_0+D\a q^4/k_F^4-i\o\right]}
\label{e1}
\ee
where $N(0)$ is the density of state, N is the orbital degeneracy,
$D=v_F^2\tau/2$ is the diffusion coefficient, $A_0=1-2t_0 N(0)$ (with
$t_0$ the base kinetic energy), $\a=3(Q/N)^2(m^8/m)^2/4$ with Q the
total charge and $k_F$ is the Fermi wave vector. For a filling $n_f$
($m/m^*=1-n_f$) close to metal-insulator phase transition $q^4$ term,
given by the quasiparticles interaction, is important and we will show
that is essential in the behavior of the electronic system.

The effect of disorder will be considered as contained in the enhancement
of the charge susceptibility and in order to analyze the effect of it on
the energy of the electronic excitations we take the general form for the
self-energy in one-loop approximation.

\section{Self-energy}

The self-energy of the electrons due to the interaction of electrons with
the charge fluctuations in the presence of disorder has the general form:
\be
\S (\bp,\o)=g^2\int \f{d^2 q}{(2\pi)^2}\int_{-\infty}^\infty \f{d\o'}{2\pi}
\left[\coth{\f{\o'}{2T}}-\tanh{\f{\tilde{\ve}(\bp+\bq)}{2T}}\right]
\f{Im\chi_c(\bq,\o')}{\o+\o'-\tilde{\ve}(\bp+\bq)+i\a}
\label{e2}
\ee
where $\tilde{\ve}(\bk)=k^2/2m-\mu$. An analytical calculation can be
performed for the two dimensional case. Using the approximation
\be
\tilde{\ve}(\bp+\bq)\cong \tilde{\ve}(\bp)+vq\cos(\theta)
\label{e3}
\ee
and the identity
\be
\lim_{\a\ra 0}\f{1}{\o-\tilde{\ve}+i\a}=\f{1}{i}\int_o^\infty dt
\exp{\left[i(\o-\tilde{\ve}+i\a)t\right]}=\f{1}{i}
\int_0^\infty dt \left[\cos{(\o-\tilde{\ve})t}+i\sin{(\o-\tilde{\ve})t}\right]
\label{e4}
\ee
we write Eq. (\ref{e2}) as
\be
\S"(\bp,\o)=-\f{g^2}{2\pi}\int_0^\infty dt \cos{(\o-\tilde{\ve})t}
\int_{-\infty}^\infty \f{d\o'}{2\pi} Im \chi_c(\bq,\o')\coth{\f{\o'}{2T}}
\int_0^{2\pi}\f{d\theta}{2\pi}\exp{\left[-ivqt\cos{\theta}\right]}
\label{e5}
\ee
\be
\S'(\bp,\o)=\f{g^2}{2\pi}\int_0^\infty dt \sin{(\o-\tilde{\ve})t}
\int_{-\infty}^\infty \f{d\o'}{2\pi} Im \chi_c(\bq,\o')\coth{\f{\o'}{2T}}
\int_0^{2\pi}\f{d\theta}{2\pi}\exp{\left[-ivqt\cos{\theta}\right]}
\label{e6}
\ee
These equations will be transformed if we perform the approximation
$\coth{\o/2T}\cong 2T/\o$ and in Eqs. (\ref{e5})-(\ref{e6}) consider
\bea
S_c&=&\int_{-\infty}^\infty\f{d\o'}{2\pi} Im \chi_c(\bq,\o')
\coth{\f{\o'}{2T}}\nonumber\\
&\cong& \f{T}{\pi}\int_{-\infty}^\infty d\o' \f{D q^2 \chi_0}
{(D A_0 q^2+D \a q^4/k_F^2)^2+\o'^2}=\f{\chi_0 k_F^2}{\a}\f{T}{\xi^{-2}+q^2}
\label{e7}
\eea
where
\bd
\chi_0=\f{N N(0)}{2}
\ed
\be
\xi^{-2}=\f{A_0 k_F^2}{\a}
\label{e8}
\ee

Using the exact formulas
\be
J_0(z)=\int_0^{2\pi}\f{d\theta}{2\pi}\exp{\left[-iz\cos(\theta)\right]}
\label{e9}
\ee
\be
K_0(kb)=\int_0^\infty dx\f{x J_0(xb)}{x^2+k^2}
\label{e10}
\ee
where $J_0(x)$ and $K_0(z)$ are the Bassel functions. The imaginary part of
the self-energy given by Eq.(\ref{e5}) has the form
\be
\S"(\bp,\o)=-\f{g^2T}{2\a\pi}\int_0^\infty dt
\cos{\left[(\o-\tilde{\ve}(\bp))t\right]}K_0(v_Ft\xi^{-1})
\label{e11}
\ee
where $v_F$ is the Fermi velocity. Performing the integral over t in
Eq. (\ref{e11}) we obtain
\be
\S"(\bp,\o)=-\f{g^2\chi_0k_F^2}{\a}\f{\pi}{2}\f{T}
{\sqrt{(\o-\tilde{\ve}(\bp))^2+(v_F\xi)^{-2}}}
\label{e12}
\ee
In the approximation $\o-\tilde{\ve}(\bp)\gg (v_F\xi)^{-1}$
from Eq. (\ref{e12}) we get
\be
\S"(\bp,\o)\cong -\f{g^2\chi_0T\xi^{-2}}{A_0(\o-\tilde{\ve}(\bp))}
\label{e13}
\ee
relation which satisfies $\S"\sim 1/\o$.

From Eq. (\ref{e13}) we can see that due to the coupling with the charge
fluctuations the electronic excitations present a non-Fermi behavior,
obtained also at $T=0$ by Wang et al. \cite{4}, but was considered as a
holon like propagation of the charge fluctuations. The result expressed by
Eq. (\ref{e13}) has been obtained in the approximation $\o\ll T$ and
following the method proposed by Vilk and Tremblay \cite{12} (See Appendix
D of Ref. \onlinecite{12} for an accurate discussion about the enhancement
of the Fermi behavior in a non-Fermi behavior of electrons interacting with
fluctuations).

An important approximation for this calculation is the existence of an
energy scale for the charge fluctuations in the presence of weak disorder.
More than that this energy scale characterized by a frequency $\o_0$ has to
satisfy the condition $\o_0\ll T$, because only the coupling of the
electrons with low energy fluctuations gives a non-Fermi behavior. In this
model we require because of the weak disorder, that $\ve_F \tau=c$ has to be
large. Then we can define $\o_0\cong\tau^{-1}=c/\ve_F$ and the condition
$\o_0\ll T$ becomes $ c\ll T E_F$.

\section{Discussion}

We showed that a non-Fermi behavior may appear by the coupling of electrons
in the presence of disorder to the charge fluctuations. Such a mechanism
was also proposed for the coupling of electrons with two-dimensional spin
fluctuations \cite{13,14}. The coupling between electrons and fluctuations
near the quantum critical point has been also proposed \cite{15,16} as the
explanation for the non-Fermi behavior of the electronic system and it seems
to be an appropriate mechanism in the heavy fermion systems.

Our model can be a good explanation for the experimental data obtained by
Boebinger et al. \cite{17} on La$_{2-x}$Sr$_x$CuO$_4$ which present a linear
dependence of the temperature at the insulator-metal crossover. Recently a
similar behavior has been observed for Pr$_{2-x}$Ce$_x$CuO$_4$ (this is an
electron-doped system)\cite{18} at low temperature.



\begin{references}
\bibitem{1} A. M. Finkelshtein, Zh. Eksp. Teor. Fiz. 84, 168 (1983) [Sov.
Phys. JETP 57, 97 (1985)]
\bibitem{2} D. Belitz and T. Kirkpatrick, Rev. Mod. Phys. 66, 261 (1994)
\bibitem{3} G. Zimanyi and E. Abrahams, Phys. Rev. Lett. 64, 2719 (1990)
\bibitem{4} Z. wang, Y. Bang and G. Kotliar, Phys. Rev. Lett. 67, 2733 (1991)
\bibitem{5} W. N. Huang and J. W. Rasul, Phys. Rev. B45, 3995 (1992)
\bibitem{6} W. N. Huang and J. W. Rasul, J. Phys. Condens Matter 5, 8877 (1993)
\bibitem{7} W. Metzner and D. Vollhardt, Phys. Rev. Lett. 62, 324 (1989)
\bibitem{8} A. Gorges and G. Kotliar, Phys. rev. B45, 6479 (1992)
\bibitem{9} V. Dobrosavljevic and G. Kotliar, Phys. Rev. B50, 1430 (1994)
\bibitem{10} A. Gorges and G. Kotliar, Rev. Mod. Phys. 68, 13 (1996)
\bibitem{11} Q. Si and G. Kotliar, Phys. Rev. B48, 13881 (1993)
\bibitem{12} Y. Vilk and A.-M. Tremblay, J. Phys. (France) 7, 1309 (1997)
\bibitem{13} A. Rosch, A. Schroder, O. Stockert and H. v. Lohneysen, Phys.
Rev. Lett. 79, 159 (1997)
\bibitem{14} O. Stockert, H. v. Lohneysen, A. Rosch, N. Pyka and M.
Loewenhaupt, Phys. Rev. Lett. 80, 5627 (1998)
\bibitem{15} S. G. Mishra and P. A. Sreeram, Phys. Rev. B57 (1998) 
\bibitem{16} P. Coleman, cond-mat 9809436
\bibitem{17} G. S. Boninger, Y. Ando, A. Passner, T. Kimura, M. Okuya,
J. Shimoyama, K. Kishio, T. Tamasaku, N. Ichikawa and S. Uchida, Phys. Rev.
Lett. 27, 5417 (1996)
\bibitem{18} P. Fournier, P. Mohanty, E. Maiser, S. Darzens, T. Venkatesan,
C. J. Loob, G. Czyek, R. A. Webb and R. L. Green, Phys. Rev. Lett. 81, 4720
(1998)
\end{references}
\end{document}